\documentclass[aps,pre,onecolumn]{revtex4}
\usepackage{amssymb}
\usepackage{eurosym}
\usepackage{epsfig}

\begin{document}

\title{Instability of solitons and collapse of acoustic waves in media with positive dispersion}
\author{E.A. Kuznetsov \/\thanks{%
kuznetso@itp.ac.ru}}
\affiliation{{\small  P.N.Lebedev Physical Institute of RAS, 53 Leninsky ave., 19991 Moscow, Russia%
\\
L.D. Landau Institute for Theoretical Physics of RAS, ul. Akademika Semenova 1A, Chernogolovka, 142432 Moscow region, 
Russia \\
 Skolkovo Institute of Science and Technology, Skolkovo, 143026 Moscow, Russia \\
 Space Research Institute of RAS, ul. Profsoyuznaya 84/32, 117997 Moscow, Russia\\
} }

\begin{abstract}
This article is a brief review of the results of studying the collapse of sound waves in media with positive dispersion, which is described in terms of the three-dimensional Kadomtsev-Petviashvili (KP) equation \cite{kadomtsev1970stability} . The KP instability of one-dimensional solitons in the long-wavelength limit is considered using the expansion for the corresponding spectral problem. It is shown that the KP instability also takes place for two-dimensional solitons in the framework of the three-dimensional KP equation with positive dispersion. According to B.B. Kadomtsev \cite{kadomtsev1976collective} this instability belongs to the self-focusing type. The nonlinear stage of this instability is a collapse. One of the collapse criteria is  the Hamiltonian unboundedness from below for a fixed momentum projection coinciding with the $L_2$-norm. This fact follows from scaling transformations, leaving this norm constant. For this reason, collapse can be represented as the process of falling a particle to the center in a self-consistent unbounded potential. It is shown that the radiation of waves from a region with a negative Hamiltonian, due to its unboundedness from below, promotes the collapse of the waves. This scenario was confirmed by numerical experiments \cite{KuznetsovMusherShafarenko1983, KuznetsovMusher1986}. Two analytical approaches to the study of collapse are presented: using the variational method and the quasiclassical approximation. In contrast to the nonlinear Schrödinger equation (NLSE) with a focusing nonlinearity, a feature of the quasiclassical approach to describing acoustic collapse is that this method is proposed for the three-dimensional KP equation as a system with hydrodynamic nonlinearity. Within the framework of the quasiclassical description, a family of self-similar collapses is found. The upper bound of this family corresponds to a strong collapse, in which the energy captures into the singularity is finite. The existence of such a regime is also confirmed based on the variational approach. The other boundary of the collapsing hierarchy coincides with the self-similar solution of the three-dimensional KP equation, which describes the fastest weak collapse. 

\end{abstract}

\maketitle



\section{Introduction}

\setcounter{equation}{0} It is now well known what role in 
nonlinear physics is played by the collapse of waves, the process of formation of a singularity in a finite time (see, for example, the review \cite{ZakharovKuznetsov2012}). If solitary waves - solitons - are characteristic of low-dimensional nonlinear wave systems, then collapses, in contrast to solitons, are typical for multidimensional  systems. The reason for this behavior, as is known, is related to the strengthening of the nonlinearity with increasing space dimension. For example, for one-dimensional nonlinear wave systems, such
as the Korteweg - de Vries (KDV) equation or the nonlinear Schrodinger equation (NLS), solitons play an essential role in their dynamics.
From the Hamiltonian point of view, solitons in these models, being
stationary points of the Hamiltonian $\mathcal{H}$ with fixed corresponding
lower positive definite integrals of motion, realize the minimum of $\mathcal{H}
$ and thus turn out to be stable in the Lyapunov sense. As a rule, the Hamiltonians for these models have two contributions corresponding to wave dispersion (positive definite) and nonlinearity. The balance of these two factors leads to the existence of solitons in such models. In contrast to Lyapunov-stable solitons, collapse occurs in wave systems with unbounded Hamiltonians. When
the Hamiltonian is an unbounded functional (from below or from above), the
collapse of waves leading to the formation of a singularity seems to be the
most probable scenario for the behavior of a nonlinear wave system. In this
case, the collapse is similar to the process of  a particle fall to the center in a self-consistent unbounded potential (see, for example, \cite{ZakharovKuznetsov1986,
KuznetsovMusher1986, ZakharovKuznetsov2012}).

The main aim of this paper is to give a brief review of results obtained in
studies of collapse for acoustic waves in media with weak positive
dispersion. Such collapse can be described in the framework of the
Kadomtsev-Petviashvili (KP-I) equation \cite{kadomtsev1970stability,
kadomtsev1976collective}. As known, positive dispersion for acoustic-type
waves takes place in the liquid helium $He^{4}$ at temperatures below the
lambda-point for some pressure region (see, e.g., \cite{Gurevich1980}).
Well- known example of positive dispersion demonstrates the Bogolyubov
spectrum for the linear waves in the gaseous Bose-Einstein condensate with
repulsion between atoms in the framework of the Gross-Pitaevskii equation 
\cite{gross1961structure, pitaevskii1961vortex}. In the strongly magnetized
plasma with $\beta =8\pi p/B^{2}\ll 1$ fast magneto-acoustic waves represent
another case of acoustic-type waves with positive dispersion which realizes
practically for the whole waves propagating angles (see \cite%
{KuznetsovMusher1986} where this question was discussed in detail).

The KP equation, as well known, represents the multidimensional
generalization of the KDV equation. Let us remind that the KDV as a 1D
equation can be easily derived from the hydrodynamic type systems for
acoustic waves with weak dispersion by means of the multi-scale expansion
(see, e.g. \cite{ZakharovKuznetsov1986a}). The difference in dispersions for
the KDV equation manifests itself in the fact that for the KDV solitons the
sign of the density deviation from the mean value is opposite to the
dispersion sign. For positive dispersion, these are the density wells that
propagate with the velocity less sound velocity  $c_{s}$. The situation for
negative dispersion is opposite: solitons are density humps moving with the
velocity greater $c_{s}$ (in particular, for gravity waves in the shallow
water limit these are water elevations). This physical property is a key
point to understand the origin of the KP instability for the 1D solitons
with respect to transverse perturbations \cite{kadomtsev1970stability}. Solitons in media with negative dispersion turn out to be stable, and with
positive dispersion, they are unstable. As was indicated  by B.B.
Kadomtsev \cite{kadomtsev1976collective}, this instability is of a
self-focusing nature.

The KP equation takes into account not only the dispersion, but also the
diffraction of acoustic beams in the direction transverse to their
propagation. The KP equation can be obtained according to the same scheme as
the KDV equation, taking into account the nonlinear renormalization of the
sound velocity, linear dispersion, and diffraction; all these factors are
considered weak in comparison with the main propagation (say, along $x $) of
acoustic beams with the mean sound velocity $c_ {s} $ and for this reason they
can be turned on perturbatively. Obviously, the propagation of a beam with $%
c_ {s} $ can be eliminated using the Galilean transform. The remaining small
terms, however, can be of the same order. The Hamiltonian structure
of the KP equation, which takes into account the
diffraction term, remains the same as in the case of KDV. The Hamiltonian $%
\mathcal{H} $ of the KP equation receives an additional quadratic
contribution due to diffraction. In the case of positive dispersion, the
quadratic part of $\mathcal{H}$ turns out to be positive definite (in the
case of negative dispersion, the quadratic contribution is not
sign-definite). Like the KDV equation, the KP equation, in addition to $%
\mathcal{H}$, preserves the $x$ -projection of the momentum $P$, which is a
positive definite quantity. The dynamics of acoustic waves with positive
dispersion, described by the KP equation, substantially depend on the space
dimension $d$.

As shown in \cite{Dryuma}, the two-dimensional KP equation belongs to
integrable models with the Lax representation, which makes it possible to
effectively investigate this model using the inverse scattering transform. The
application of this approach made it possible to analytically find a
localized two-dimensional soliton solution, the so-called lump \cite%
{manakov1977twodimensional}. It is necessary to mention that this solution
was found first numerically by Petviashvili \cite{Petviashvili} by the
scheme named later the Petviashvili scheme. In the form of a lump soliton,
moving along the $x$ axis at a constant velocity, it is a stationary point of
the Hamiltonian at a fixed momentum $P$. As shown by \cite%
{KuznetsovTuritsyn1982}, for a fixed $P$ the Hamiltonian turns out to be a
functional bounded from below, reaching its minimum corresponding to the
soliton solution. For this reason, the two-dimensional soliton solution
turns out to be  stable in the Lyapunov sense at $d=2$.

However, in three-dimensional geometry the situation changes radically: the
KP Hamiltonian turns out to be a functional unbounded from below (see \cite%
{KuznetsovMusherShafarenko1983, KuznetsovMusher1986}), that is easy verified
by applying two independent scaling transformations along the $x$ axis and
in the transverse plane, for each of which $P$ remains invariant. It is this
property that became the starting point for numerical experiments to observe
the collapse \cite{KuznetsovMusherShafarenko1983} (see also \cite%
{KuznetsovMusher1986}).

The results of these numerical experiments showed a tendency towards the formation of a singularity in a finite time. In these calculations, the influence of low-amplitude wave radiation on the collapse process was studied. It was shown that radiation contributes to the collapse leading to the formation of weak-type singularities, when the wave energy trapped in the singularity formally tends to zero, but in reality, due to small dissipation, it turns out to be a finite but rather small value. As was shown in \cite{KuznetsovMusher1986}, for a strongly magnetized plasma, the collapse of fast magnetoacoustic waves
 plays a very significant role in the dynamics of oblique (relative to the average magnetic field) shock waves, determining the fine structure of the front and its width. The collapse in this case is an efficient mechanism for the transfer of wave energy to ions.

It should be noted that approximately at the same time when numerical experiments were carried out to observe the collapse of sound waves, i.e. around 1985, a three-dimensional quasi-classical theory of wave collapse was created within the framework of the nonlinear Schrodinger equation (NLSE)
\cite{ZakharovKuznetsov1986}. In this work, a classification of wave collapses was given, starting with a strong collapse, when the energy captured into the singularity is finite, and ending with weak collapses. In \cite{ZakharovKuznetsov1986} (see also the review \cite{ZakharovKuznetsov2012}) it was established that the strong collapse solution for 3D NLSE corresponds to the quasiclassical compression of the wave packet as a whole. In addition, the existence of a whole family of quasiclassical weak collapses was pointed out. The upper bound of this family coincides with the self-similar solution, which describes the fastest process of singularity formation in the weak collapse regime.

This review presents a quasiclassical theory of acoustic collapse following \cite{BlahaKuznetsovLaedkeSpatschek}. This kind of description of collapses as applied to acoustic waves with positive dispersion has a number of features related to the non-linearity of the hydrodynamic type, which is different from that in NLSE. Despite this, it is possible to develop a variational approach for the three-dimensional KP equation. As is well known, such an approach can claim only some qualitative description of the collapse. It is important to note that the solution within the framework of the variational approach to NLSE gave the same self-similarity as that for the exact quasiclassical solution found in \cite{ZakharovKuznetsov1986}.
The same behavior is observed with respect to
 acoustic collapse in the quasiclassical limit. For the three-dimensional KP equation, as well as for the 3D NLSE, self-similarity is found for a whole family of quasiclassical collapsing distributions.

It should be noted that within the framework of the three-dimensional KP equation, one can see much in common in the problem of the collapse of the KDV equation with power nonlinearity ($\sim u^{p-1}u_{x}$), another example of a system with hydrodynamic nonlinearity \cite{BlahaLaedkeSpatschek}. This equation
exhibits critical behavior at exponent $p=6$. For a smaller value of $p$, the Hamiltonian turns out to be bounded from below for a fixed momentum $P$. In this case, the solitons realize the minimum of $\mathcal{H}$ and, accordingly, are Lyapunov stable \cite{Kuznetsov1984} (see also \cite{ZakharovKuznetsov2012}). For $p\geq 6$, the Hamiltonian becomes an unbounded functional, and therefore a collapse occurs in this case, which was observed numerically in \cite{BlahaLaedkeSpatschek}. Note that for the collapse in the KDV equation with a power nonlinearity, as well as for the three-dimensional KP equation, self-similarity is found for a whole family of quasiclassical collapsing distributions, there is a correspondence between the quasiclassical self-similar solution and the predictions of the variational approach. 

The outline of this article is as follows. Section II discusses
general properties of the KP equation for media with positive dispersion.
In this section, as an example, we first present the derivation of the KP equation based on the defocusing NLSE by means of of multi-scale expansion (for this method, see, for example, \cite{ZakharovKuznetsov1986a}).  In this case, a one-dimensional soliton (a soliton solution of the KDV equation) is a density well propagating at a velocity less than the average velocity of sound. It is this property that is the key to understanding the nature of KP instability \cite{kadomtsev1970stability, kadomtsev1976collective}. Next, we discuss the Hamiltonian structure of the KP equation and the general properties of solitons as stationary points of the Hamiltonian at a fixed momentum $P$. It is found how the value of the Hamiltonian on soliton solutions for an arbitrary dimension $d$ is expressed in terms of the momentum of the solitons. These relations are obtained by scaling transformations that preserve $P$.
Section III is devoted to the KP instability of solitons with respect to long-wavelength transverse perturbations. First, we consider the problem of the stability of one-dimensional (KDV) solitons - the key one, in our opinion, following the method proposed in \cite{KuznetsovTuritsyn1982}. Further, this approach is used in the analysis of the instability of two-dimensional KP solitons with respect to three-dimensional perturbations (analogous to the three-dimensional KP instability). Based on this method, the KP-type instability for gray/dark solitons was predicted in the framework of the defocusing NLSE \cite{KuznetsovTuritsyn1988}. The existence of this instability was confirmed experimentally, as well as numerically in \cite{Schwartzlander, Law} for stationary propagation of electromagnetic waves in defocusing Kerr media. In these works, it was found that the nonlinear stage of this instability leads to the formation of a periodic chain of point vortices.  

In the same section, we consider the  stability problem for two-dimensional solitons for a defocusing NLSE. Solitons of this type have a number of features. When the depth
wells of soliton density is sufficiently small, such solitons asymptotically coincide with lumps. With increasing depth in the soliton and, respectively,
decrease in its velocity $V$, as was first shown numerically by \cite{Jones}, there is some
the critical value $V_{cr}$, when the density at the center of the soliton vanishes. Below this critical value, the soliton solution undergoes a bifurcation associated with the appearance of two zeros in the density corresponding to
two vortices with opposite circulations. This is a dipole pair propagating in a direction perpendicular to the dipole. With a further decrease in the
soliton velocity, this pair turns into a vortex pair for a two-dimensional
Euler hydrodynamics. As shown in \cite{KuznetsovRasmussen}, this entire family of two-dimensional solitons is unstable with respect to transverse perturbations (along the third axis $z$). For a small soliton amplitude, this instability is analogous to the KP instability for a lump; for solitons propagating with velocities less than critical, this instability transforms into the Crow instability, which was first discovered for a dipole vortex pair in terms of the incompressible Euler equations \cite{Crow}. 

 The fourth section deals with collapse for the three-dimensional KP equation. It is shown that the Hamiltonian $\mathcal{H}$ is unbounded from below, which follows from scaling transformations that preserve $P$. Due to the unboundedness of the Hamiltonian, the radiation of small-amplitude waves in their region with a negative value of $\mathcal{H}$ promotes the collapse, due to which weak-type singularities are formed. Further, two analytical approaches to the study of collapse are presented: using the variational method and the semiclassical approximation. It is found that the solution of the quasiclassical Whitham equations for strong collapse and ansatz in the framework
variational approach have the same self-similar asymptotics near the singularity. Within the framework of the quasiclassical description, a family of self-similar collapses is found. The upper bound of this family corresponds to a strong collapse, in which the energy captured into the singularity is finite. The existence of such a regime is also confirmed based on the variational approach. The other boundary of the collapsing hierarchy coincides with the self-similar solution of the three-dimensional KP equation, which describes the fastest weak collapse.

In conclusion, we list the main results of this review and discuss problems that have not yet been solved. Among these problems, the most important one concerns finding a sufficient condition for collapse. Recall that for the focusing NLSE, such a criterion was first obtained in the two-dimensional case by Vlasov, Petrishchev, and Talanov \cite{VlasovPetrishchevTalanov}, and then generalized by Zakharov to
three-dimensional NLSE \cite{Zakharov1972}. The sufficient condition coincides with the negativeness of the
Hamiltonian $\mathcal{H}$ for both $d=2$ and $d=3$.

\section{General properties of the KP equation and soliton solutions}

\setcounter{equation}{0}
Consider first the general properties of the KP equation with positive dispersion,
often called the KP-I equation.
First of all, we will show how this equation is derived from the equation of motion for media with hydrodynamic nonlinearity. As an example, consider the defocusing nonlinear Schrodinger equation
\begin{equation}
i\frac{\partial \psi }{\partial t}+\frac{1}{2}\Delta \psi +(1-|\psi
|^{2})\psi =0. \label{NLS}
\end{equation}%
This equation is written in dimensionless form. As applied to gaseous Bose-Einstein condensates (BEC) at a temperature $T\to 0$ $\psi $ has the meaning of the wave function of the condensate (in this context, the equation (\ref{NLS}) is called the Gross-Pitaevskii equation), $|\psi |^{2}$ is the density of the condensate $n$ . In a non-linear
optics $\psi $ is the envelope of the wave packet of the electric field in the media
with defocusing Kerr nonlinearity.

We will assume that the wave function tends to
to $1$ as $r\rightarrow \infty $, and accordingly the density is $n\rightarrow 1$. In that
case, as is well known, small-amplitude waves have a dispersion law
\begin{equation}
\omega =k\sqrt{1+k^{2}/4}, \label{Bogolyubov}
\end{equation}
the so-called Bogolyubov spectrum. For $k\rightarrow 0$
 this spectrum
transforms into an acoustic-type dispersion law with positive dispersion
\begin{equation}
\omega =k(1+k^{2}/8+...), \label{linear}
\end{equation}%
where the sound velocity $c_{s}=1$. In order to derive the KP equation from 
\ref{NLS}) consider a beam of acoustic waves of small (but finite) amplitude propagating mainly in one direction (say, along $x$) with a small angular distribution
($k_{\perp }/k_{x}\ll 1$) . The latter means that for such wave beams the dispersion relation (\ref{linear}) is written as 
\begin{equation}
\omega =k_{x}\left[ 1+k_{x}^{2}/8+\frac{1}{2}\left( k_{\perp }/k_{x}\right)
^{2}\right] .  \label{small}
\end{equation}
In such a case the first term in the bracket can be excluded by the Galilean
transformation to the system of coordinates moving with sound velocity along 
$x$-direction so that (\ref{small}) has the form of the standard KP dispersion
relation%
\begin{equation}
\Omega =k_{x}\left[ k_{x}^{2}/8+\frac{1}{2}\left( k_{\perp }/k_{x}\right)
^{2}\right]  \label{KPdispersion}
\end{equation}%
which takes into account both weak positive dispersion and diffraction in a
transverse direction. Two terms in (\ref{KPdispersion}) become comparable if 
$k_{x}^{4}\propto k_{\perp }^{2}$. 
It is this simple observation that is one of the key ones in the derivation of the KP equation.

Next, we rewrite the NLSE as a system of hydrodynamic type for density $n$ and phase $\varphi
$ ( $\psi =\sqrt{n}e^{i\varphi }$):%
\begin{eqnarray*}
\frac{\partial n}{\partial t}+\nabla (n\nabla \varphi ) &=&0, \\
\frac{\partial \varphi }{\partial t}+\frac{1}{2}(\nabla \varphi )^{2}+n-1 &=&%
\frac{\Delta \sqrt{n}}{2\sqrt{n}},
\end{eqnarray*}
To derive the KP equation, we rewrite the NSE as a system of hydrodynamic type for density $n$ and phase $\varphi
$ ( $\psi =\sqrt{n}e^{i\varphi }$):%
\begin{eqnarray*}
\frac{\partial n}{\partial t}+\nabla (n\nabla \varphi ) &=&0, \\
\frac{\partial \varphi }{\partial t}+\frac{1}{2}(\nabla \varphi )^{2}+n-1 &=&%
\frac{\Delta \sqrt{n}}{2\sqrt{n}}.
\end{eqnarray*}%
Further, we will apply a multiscale expansion to this system, following the work \cite%
{ZakharovKuznetsov1986a} (see also\cite{KuznetsovTuritsyn1988}).

We introduce slow time and slow coordinates,
\[
t^{\prime }=\varepsilon ^{3}t,x^{\prime }=\varepsilon (x-t),r_{\perp
}^{\prime }=\varepsilon ^{2}r_{\perp },
\]%
and we will look for a solution to this system in the form of a series in powers of a small parameter
$\varepsilon $,%
\[
n=1+\sum_{k=1}^{\infty }\varepsilon ^{2k}n_{k}(x^{\prime },r_{\perp
}^{\prime },t^{\prime }).
\]%
The first two orders are easily found
(see, for example, \cite{ZakharovKuznetsov1986a, KuznetsovTuritsyn1988} ).
The KP-I equation is obtained in the framework of this perturbation theory as a third-order solvability condition ($\sim\varepsilon ^{3}$):
\begin{equation}
\frac{\partial }{\partial x}\left( n_{1t}+\frac{3}{2}n_{1}n_{1x}-\frac{1}{8}%
n_{1xxx}\right) =-\frac{1}{2}\Delta _{\perp }n_{1} \label{NLS-KP}
\end{equation}%
where $\Delta _{\perp }=\partial _{y}^{2}+\partial _{z}^{2}$ is the transverse Laplace operator (primes omitted below).

As is known, the KP equation (\ref{NLS-KP})
is a multidimensional generalization of the one-dimensional KDV equation ($\Delta _{\perp}=0$). In the latter case, the soliton solution has the form
\begin{equation}
n_{1}=-\frac{\nu ^{2}}{\cosh ^{2}\nu (x+\nu ^{2}t/2-x_{0})}
\label{KDVsoliton}
\end{equation}%
which are density wells ($n_1 <0$) moving at $c_{s}-\nu ^{2}/2$, less than the acoustic speed. Note that this condition also holds for multidimensional solitons. This requirement is equivalent to the absence of a Cherenkov resonance between a soliton and linear waves (see, for example, \cite{ZakharovKuznetsov1998}).  As a result, therefore, in the 1D case, the soliton velocity decreases with increasing soliton amplitude. This property is key for understanding the origin of the KP instability \cite{kadomtsev1970stability}  with respect to perturbations transverse to the KDV soliton. As shown by B.B. Kadomtsev \cite{kadomtsev1976collective}, this instability has
 self-focusing character. Let the soliton be modulated in the transverse direction. Then, according to (\ref{KDVsoliton}), regions of the soliton with a larger amplitude will move more slowly than those regions that have a lower amplitude. As a result, regions with a lower amplitude will overtake regions with an increased amplitude, which will lead to focusing of the soliton front and accumulation of its energy. It is easy to see that these arguments also work at the non-linear stage of instability, i.e. at the stage of collapse formation, described in terms of the three-dimensional KP equation (\ref{KDVsoliton}) (see the following sections).

\subsection{Hamiltonian structure}

Before discussing the Hamiltonian structure of the KP equation, solitons and their stability, we rewrite (\ref{NLS-KP}) in the standard form, which can be obtained using simple substitutions and rescaling:
\begin{equation}
\frac{\partial }{\partial x}\left( u_{t}+6uu_{x}+u_{xxx}\right) =\Delta
_{\perp}u. \label{KP-stand}
\end{equation}
In the one-dimensional case, this equation goes into the standard form adopted for
KDV. Like KDV, the equation (\ref{KP-stand}) belongs to the Hamiltonian equations
\[
u_{t}=\frac{\partial }{\partial x}\frac{\delta \mathcal{H}}{\delta u}
\]%
with the Hamiltonian
\begin{equation}
\mathcal{H}=\int \left[ \frac{u_{x}^{2}}{2}+\frac{\left( \nabla _{\perp
}w\right) ^{2}}{2}-u^{3}\right] dr\equiv \frac{1}{2}I_{1}+\frac{1}{2}%
I_{2}-I_{3}, \label{H-KP}
\end{equation}%
where $w_{x}=u$. Here the first term is responsible for the wave dispersion,
the second one takes into account the diffraction of an acoustic beam in the transverse direction, the last term in $\mathcal{H}$ describes
nonlinearity of acoustic waves. Besides $\mathcal{H}$, the KP equation preserves the total momentum, $x-$ component of which is a positive definite quantity, $P=\frac{1}{2}\int u^{2}dr$. The KDV equation follows from this for $w=0$. In variables $u$ the KDV soliton (\ref{KDVsoliton}) is rewritten as %
\begin{equation}
u_{s}=\frac{2\kappa ^{2}}{\cosh ^{2}\kappa \left( x-4\kappa
^{2}t-x_{0}\right) }. \label{KDV}
\end{equation}%
which is a stationary point of the Hamiltonian $\mathcal{H}$ for a fixed $P$,
\begin{equation}
\delta (\mathcal{H}+VP)=0. \label{stat-point}
\end{equation}
This variational problem is equivalent to the stationary KDV equation
\begin{equation}
-Vu_{x}+6uu_{x}+u_{xxx}=0 \label{stat}
\end{equation}%
with a soliton solution (\ref{KDV}) moving at $V=4\kappa
^{2}$. In the multidimensional case, namely, within the framework of the KP equation (\ref{KP-stand}), soliton solutions propagating along the $x$ axis with constant velocities $V$ are also stationary points of $\mathcal{ H}$ for a fixed $ P$. In the two-dimensional case, these are the so-called lump solutions localized in all directions with power-law asymptotics as $r\rightarrow \infty $. For the first time such solutions were discovered numerically by Petviashvili \cite{Petviashvili}. Then at work
\cite{manakov1977twodimensional} lumps were found as exact solutions of the KP equation based on the application of the inverse scattering transform.

In the three-dimensional case, solitons of this type can be found numerically, and possibly also analytically, using the Hirota  scheme \cite{Hirota}. The Hirota representation, as was found relatively recently in a number of papers (see \cite{Ma}-\cite{Kaptsov2021}), is applicable to the KP equation for arbitrary dimension. However, this question - the question of analytically finding a soliton solution in three dimensions - remains open. 

Despite this, for solitons of the KP equation (\ref{KP-stand}) with an arbitrary space dimension $d$, one can establish a number of relations, in particular, the relationship between the values of the Hamiltonian and the momentum $P$ on the soliton solution. These relations are consequences of the variational problem (\ref{H-KP}), which is valid for an arbitrary dimension $d$.
This fact was first established in \cite{KuznetsovTuritsyn1982} (see also \cite{KuznetsovMusherShafarenko1983, KuznetsovMusher1986}). To do this, we need to find relations between the integrals in $\mathcal{H}$. 
The first relation is obtained if the equation for soliton,%
\[
-Vu_{x}+6uu_{x}+u_{xxx}=\Delta _{\perp }w  
\]
is first multiplied by $w$ and then integrated over $r$. It gives%
\begin{equation}
2VP+I_{1}+I_{2}-3I_{3}=0.  \label{1st}
\end{equation}

The other two relations follow from the variational problem (\ref{stat-point})
after applying  scaling transformations preserving  $P$,
\[
u(x,r_{\perp })\rightarrow a^{-1/2}b^{(1-d)/2}u(x/a,r_{\perp }/b).
\]%
Under these transformations, $\mathcal{H}$ becomes a function of scaling
parameters $a$ and $b$:%
\begin{equation}
\mathcal{H}(a,b) =\frac{1}{2}a^{-2}I_{1}+\frac{1}{2}%
a^{2}b^{-2}I_{2}-a^{-1/2}b^{(1-d)/2}I_{3}. \label{H-scaling}
\end{equation}%
According to the variational problem (\ref{stat}) on the soliton solution, the derivatives of $\mathcal{H}(a,b) $
with respect to $a ,b$ must vanish at $a =b =1$, which
leads to two additional relations: %
\begin{eqnarray}
-I_{1}+I_{2}+\frac{1}{2}I_{3} &=&0, \label{2nd} \\
-I_{2}-\frac{1-d}{2}I_{3} &=&0. \label{3d}
\end{eqnarray}%
Further expressing $I_{1}$, $I_{2}$ and $I_{3}$ from the relations (\ref{1st},\ref{2nd},\ref{3d}) in terms of $P_{s}$ on the soliton solution, for $\mathcal{H}_{s}$ we find:
\[
\mathcal{H}_{s}=\frac{2d-5}{7-2d}VP_{s}.
\]
In the one-dimensional and two-dimensional cases, the Hamiltonian on the soliton solution is negative, equal to $-\frac{3}{5}VP_{s}$ and $-\frac{1}{3} 
VP_{s}$ respectively. Moreover, in both cases the Hamiltonian turns out to be bounded from below for a fixed $P$  not only under scaling transformations, but also under general perturbations \cite{KuznetsovTuritsyn1982} (see also \cite{ZakharovKuznetsov2012}). In this case, $\mathcal{H}$ reaches its minimum on the soliton solution in the one-dimensional and two-dimensional cases. It means,
that the solitons for $d=1$ and $2$ are Lyapunov stable.

In three dimensions, the situation is opposite: the Hamiltonian on the soliton $\mathcal{H}%
_{s}=$ $VP_{s}$ is positive, and the corresponding stationary point is
saddle (for this reason, the three-dimensional soliton is unstable, see Section IV) and, in addition, $\mathcal{H}$ turns out to be a functional that is unbounded from below. To verify the latter, it suffices to consider the lines of steepest descent $a^{2}/b=const$ for
function $\mathcal{H}( a,b) $.
The unbounded nature of the Hamiltonian, due to the nonlinear term, leads to the fact that with decreasing scales $a,b$ its role becomes dominant in comparison with both the dispersion and diffraction terms, while $\mathcal{H}\to -\infty $. As we will see in the following sections, the unboundedness of the Hamiltonian from below plays a very important role in the formation of the singularity, i.e., in the collapse.

\section{KP instability}

The KP instability for the KDV solitons was investigated first time by
Kadomtsev and Petviashvili \cite{kadomtsev1970stability,
kadomtsev1976collective} in the long-wave limit. Here we will apply the
approach different from that used by Kadomtsev and Petviashvili. It will be
shown how the KP instability of the 1D soliton can be analyzed in the
long-wave limit by means of the perturbation technique following \cite%
{KuznetsovTuritsyn1982} (see also \cite{KuznetsovRubenchikZakharov, KivsharPelinovsky}). It is worth noting that originally the method of
this paper was applied to demonstrate the instability of 2D solitons
relative to three-dimensional perturbations. Later it was successfully used
for stability problems of the 1D dark/gray solitons in the framework of the
defocusing NLSE relative to transverse perturbations \cite%
{KuznetsovTuritsyn1988}, as well as of the 2D solitons against 3D
perturbations \cite{KuznetsovRasmussen}. Then this approach demonstrated its efficiency for 1D solitons propagating in the boundary layer \cite%
{DyachenkoKuznetsov} in the framework of the Shrira equation \cite{Shrira}
(2D generalization of the Benjamin-Ono equation).

\subsection{Instability of KDV solitons}

We start our analysis from stability of the one-dimensional solitons.

Solution in the KP equation (\ref{KP-stand}) will be sought in the form%
\[
u=u_{s}(x-Vt)+\psi (x-Vt,r_{\perp },t), 
\]%
where $\psi $ is assumed to be small perturbation in comparison with the KDV
soliton $u_{s}$ (\ref{KDV}).

Then by translating to the system moving with velocity $V$ and linearizing (%
\ref{KP-stand}) on the background $u_{s}$, for perturbation $\psi
(x,r_{\perp },t)=\varphi (x)\exp \left( -i\omega t+ikr_{\perp }\right) $ we
arrive at the following spectral problem,%
\[
-i\omega \frac{\partial }{\partial x}\varphi +k^{2}\varphi =-\frac{\partial
^{2}}{\partial x^{2}}\left( -V+6u_{s}+\frac{\partial ^{2}}{\partial x^{2}}%
\right) \varphi \equiv L\varphi .  
\]

We will seek for solution of this equation in the long-wave limit, expanding 
$\varphi $ and $\omega $ in series relative to small $k$:%
\begin{eqnarray*}
\varphi &=&\varphi _{0}+\varphi _{1}+\varphi _{2}+..., \\
\omega &=&\omega _{1}+\omega _{2}+...
\end{eqnarray*}
In the leading order, 
\[
-\frac{\partial ^{2}}{\partial x^{2}}\left( -V+6u_{s}+\frac{\partial ^{2}}{%
\partial x^{2}}\right) \varphi _{0}\equiv L\varphi _{0}=0. 
\]%
Comparing with (\ref{stat}) it is easy to see that this equation is
satisfied for $\varphi _{0}=\frac{\partial }{\partial x}u_{s}$. This
neutrally stable mode corresponds to a small shift of the soliton as a whole.

In the next order we have%
\[
-i\omega _{1}\frac{\partial }{\partial x}\varphi _{0}=L\varphi _{1}. 
\]%

In order to find $\varphi _{1}$ , consider equation (\ref{stat}) 
derivative of which with respect to $V$  is written in the form
\[
-\frac{\partial u_{s}}{\partial x}=-\frac{\partial }{\partial x}\left(
-V+6u_{s}+\frac{\partial ^{2}}{\partial x^{2}}\right) \frac{\partial u_{s}}{%
\partial V}. 
\]
Hence, from comparison of two last equations we conclude that%
\[
\varphi _{1}=i\omega _{1}\frac{\partial u_{s}}{\partial V}. 
\]%
Thus, the second order reads as follows%
\begin{equation}
-i\omega _{1}\frac{\partial }{\partial x}\varphi _{1}+k^{2}\varphi
_{0}=L\varphi _{2}.  \label{second}
\end{equation}

The solvability condition of this equation (for $\varphi _{2}$) consists in
the orthogonality the left hand side of this equation to the eigen function $%
\chi $ with zero eigen value of the operator conjugated to $L,$%
\[
\left( -V+6u_{s}+\frac{\partial ^{2}}{\partial x^{2}}\right) \frac{\partial
^{2}}{\partial x^{2}}\chi =0. 
\]%
It is easy to see that 
\[
\frac{\partial }{\partial x}\chi =u_{s}, 
\]
i.e. $\chi $ is the hydrodynamic potential $w_{s}$ for the soliton. Hence, substituting $\varphi
_{0}=\frac{\partial }{\partial x}u_{s}$ and $\varphi _{1}=i\omega _{1}\frac{%
\partial u_{s}}{\partial V}$ into (\ref{second}) and then scallarily
multiplying by$\chi $ we get%
\[
\omega _{1}^{2}\int \chi \frac{\partial }{\partial x}\frac{\partial u_{s}}{%
\partial V}dx+k^{2}\int \chi \frac{\partial }{\partial x}u_{s}=0. 
\]
Integration by parts in both integrals gives 
\[
\omega ^{2}=-k^{2}\frac{2P}{\partial P/\partial V}. 
\]%
For the KDV soliton the momentum $P\varpropto V^{3/2}$ and finally we have 
\begin{equation} \label{KP-KDV}
\omega ^{2}=-\frac{4V}{3}k^{2}<0. 
\end{equation}
Thus, we arrive at instability. This is precisely the Kadomtsev-Petviashvili
instability. It is worth noting that the exact expression of the growth rate
of this instability for 1D soliton was found by Zakharov by means of the
inverse scattering transform (IST) applying to the KP equation (\ref%
{KP-stand}).

The reason for this instability, as explained by B.B. Kadomtsev in his book \cite{kadomtsev1976collective}, is due to the fact that solitons of acoustic waves
with positive dispersion represent density wells whose velocity decreases with increasing soliton amplitude (see, for example, the equation (\ref{KDVsoliton})). As a consequence, if such a soliton is modulated in the transverse
 direction, then areas of higher amplitude will lag behind areas of lower amplitude. This leads to an instability of the self-focusing type, the development of which leads to focusing of the soliton front and accumulation of wave energy.

These arguments are valid for all the KP-type instabilities mentioned above, namely, the instabilities of one-dimensional dark/gray solitons in the framework
defocusing NLSE with respect to transverse perturbations \cite%
{KuznetsovTuritsyn1988}, two-dimensional solitons against the three-dimensional perturbations \cite{KuznetsovRasmussen}, as well as for one-dimensional solitons propagating in the boundary layer \cite{DyachenkoKuznetsov}, within the so-called Shrira equation \cite{Shrira}. Since the KP instability belongs to the
self-focusing type, such argumentation can be also applied to the nonlinear
stage of such instability in the 3D KP equation with positive dispersion.
It should be noted that in the two-dimensional case, the presence of stable two-dimensional solitons - lumps significantly affects the development of the KP instability. As shown in \cite{PelinovskyStepanyants}, instability at the nonlinear stage leads to the detachment of a chain of lumps from the soliton, which, in turn, also experiences instability with respect to transverse perturbations.

\subsection{Instability of two-dimensional KP solitons}

Consider now, following to \cite{KuznetsovTuritsyn1982}, the linear stability
problem for lumps relative  to perturbations along the third coordinate $z$ for the
lumps. The solution of this type $u=u_{s}(x-Vt,y)$ propagating with the
velocity $V$ along the $x$-axis satisfies the stationary KP equation, 
\begin{equation}
\left[ -V\frac{\partial ^{2}}{\partial x^{2}}+\frac{\partial ^{4}}{\partial
x^{4}}-\frac{\partial ^{2}}{\partial y^{2}}\right] u_{s}+\frac{\partial ^{2}%
}{\partial x^{2}}u_{s}^{2}.  \label{stat-2D}
\end{equation}%
First time solution of this equation was found numerically by V.I. Petviashvili \cite{Petviashvili}. Later on, the exact solution in the form of the lump was obtained in \cite{manakov1977twodimensional} by means of the inverse scattering transform.

To solve the linear stability problem for the lump, we assume
the perturbation is small, depending on time and coordinate $z$ exponentially: $\psi =\psi (x-Vt,y)\exp {-i\omega t+ikz}$. Then the linearization (\ref{KP-stand}) against the background of the stationary solution of the equation (\ref{KP-stand}) leads to the following spectral problem: 
\begin{equation}
A\psi -i\omega \frac{\partial \psi }{\partial x}=-k^{2}\psi ,
\label{2DKP-stability}
\end{equation}%
where the operator 
\[
A=\frac{\partial ^{2}}{\partial x^{2}}[-V+6u_{s}]+\frac{\partial ^{4}}{%
\partial x^{4}}-\frac{\partial ^{2}}{\partial y^{2}}.  
\]
As in the previous example, the solution of this eigenvalue problem will be
sought in the long-wave limit, $k\rightarrow 0$, in the form of series, 
\[
\psi =\psi _{0}+\psi _{1}+\psi _{2}+...
\]%
Here $\psi _{0}$ is determined by the equation 
\[
A\psi _{0}=0,
\]%
namely, $\psi _{0}$ is a neutrally stable mode which can be represented as
the infinitesimal translations of the soliton as a whole along $x$ and $y$
directions, 
\[
\psi _{0}=C_{1}\frac{\partial u_{s}}{\partial x}+C_{2}\frac{\partial u_{s}}{%
\partial y}.
\]%
Both shifts, evidently, are independent and can be considered separately. In
the next order for the first mode we have, 
\[
A\psi _{1}-i\omega \frac{\partial }{\partial x}u_{sx}=0.
\]%
As in the 1D case, differentiation of the stationary equation (\ref{stat-2D}%
) with respect to $V$ allows to find  $\psi _{1}$:
\[
\psi _{1}=i\omega \frac{\partial u_{s}}{\partial V}.
\]%
The second approximation reads 
\begin{equation}
A\psi _{2}=-\omega ^{2}\frac{\partial ^{2}u_{s}}{\partial x\partial V}-k^{2}%
\frac{\partial u_{s}}{\partial x}.  \label{second-1}
\end{equation}%
To obtain $\omega (k)$ from this equation one needs to use the solvability
condition that is orthogonality of the right-hand side of this equation to
the eigenfunction $\phi _{0}$ with zero-eigenvalue for the operator adjoint
of $A$. It is easy to find that $\phi _{0}=w_{s}$. Hence we get 
\[
\omega ^{2}\int w_{s}\frac{\partial ^{2}u_{s}}{\partial x\partial V}%
dxdy=-k^{2}\int w_{s}\frac{\partial u_{s}}{\partial x}dxdy.
\]%
Integrating by parts yields 
\[
\omega ^{2}\frac{\partial P}{\partial V}=-2k^{2}P.
\]%
For the lump solution $P\sim V^{1/2}$ and as a result we get instability 
\[
\omega ^{2}=-4k^{2}V<0.
\]%
For the neutral mode $\frac{\partial u_{s}}{\partial y}$ such a procedure
leads to stability.

\subsection{KP instability of gray/dark solitons}

For the NLS equation (\ref{NLS}) a solution in the form of gray/dark
solitons $\psi _{0}(x-\kappa t))$ is found in the form of a solitary
one-dimensional wave propagating with velocity $\kappa $ . In this case, $%
\psi _{0}(x-\kappa t))$ satisfies the following ordinary differential
equation, 
\begin{equation}
-i\kappa \frac{\partial \psi _{0}}{\partial x}+\frac{1}{2}\frac{\partial
^{2}\psi _{0}}{\partial x^{2}}+(1-|\psi _{0}|^{2})\psi _{0}=0,\,\,|\psi
_{0}|^{2}\rightarrow 1,\,\,\mbox{as}\,\,|x|\rightarrow \infty .
\label{statNLS}
\end{equation}%
Its solution is written as 
\begin{equation}
\psi _{0}=\nu \tanh \nu (x-\kappa t-x_{0})+i\kappa ,\,\,\kappa ^{2}+\nu
^{2}=1.  \label{gray}
\end{equation}%
When $\kappa =0$ this solution presents a dark soliton (or a domain wall).
If $\kappa \neq 0$ we have gray solitons. When $\kappa \rightarrow 1=$ to
sound velocity, these solitons transform into the KDV solitons (\ref%
{KDVsoliton}).

To consider the stability problem of such solutions we will follow to \cite%
{KuznetsovTuritsyn1988}. Assuming small perturbations $\phi _{1}=\delta \psi
,\,\phi _{2}=\delta \psi ^{\ast }$ propagating with the velocity $\kappa $
and depending on time and $y$ exponentially, $\propto \exp (-i\omega t+iky)$%
, we arrive at the following eigen value problem for the matrix differential
operator, 
\begin{equation}
\omega \sigma _{3}u-\frac{1}{2}k^{2}u+Lu=0,  \label{gray-stability}
\end{equation}%
where $\sigma _{3}=\left[ 
\begin{array}{cc}
1 & 0 \\ 
0 & -1%
\end{array}%
\right] $ is the Pauli matrix, $u=\left( 
\begin{array}{c}
\phi _{1} \\ 
\phi _{2}%
\end{array}%
\right) $, and%
\[
L=-i\kappa \sigma _{3}\frac{\partial }{\partial x}+\frac{1}{2}\frac{\partial
^{2}}{\partial x^{2}}-\left( 
\begin{array}{cc}
2|\psi _{0}|^{2}-1 & \psi _{0}^{2} \\ 
\psi _{0}^{\ast 2} & 2|\psi _{0}|^{2}-1%
\end{array}%
\right) 
\]%
is the Hermitian operator.

Solution of this eigen value problem will be sought, as in the previous case, in the long-wave limit ($%
k\rightarrow 0$) in the form of a power series with respect to $k$,%
\[
u=u_{0}+u_{1}+u_{2}+...,\,\, \omega =\omega _{1}+\omega _{2}+... 
\]%
Here $u_{0}$ is a neutrally stable mode, $Lu_{0}=0,$ corresponding to
infinitesimal translation of the soliton as a whole, 
\[
u_{0}=\frac{\partial }{\partial x}\left( 
\begin{array}{c}
\psi _{0} \\ 
\psi _{0}^{\ast }%
\end{array}%
\right) . 
\]%
In the next order, we have%
\[
\omega \sigma _{3}u_{0}+Lu_{1}=0. 
\]
As far as $u_{1}$ concerns, it is found by means of the same trick as we
used in the previous subsection. We differentiate equation (\ref{statNLS})
relative to the velocity $\kappa $ that gives 
\begin{equation} \label{1storder}
u_{1}=i\omega \frac{\partial }{\partial \kappa }\left( 
\begin{array}{c}
\psi _{0} \\ 
\psi _{0}^{\ast }%
\end{array}%
\right) . 
\end{equation}
If we substitute (\ref{gray}) into this expression, then
it turns out that $u_{1}$ tends to a constant value as $|x|\rightarrow
\infty$. At first glance, therefore, it seems that $u_{1}$ belongs to a continuous
spectrum and cannot ensure instability, since perturbations from the continuous spectrum are obviously stable. In fact, as shown in
\cite{KuznetsovTuritsyn1988}, in this case it is necessary to take into account
 a more accurate asymptotic behavior of this function at infinity based on
exact linearized equation (\ref{gray-stability}) by solving the asymptotic matching problem. This analysis gives (see \cite{KuznetsovTuritsyn1988}) that $u_{1}$ at $|x|\rightarrow \infty $
vanishes with exponent $\sim k$ and, for this reason, belongs to a bound
state.

The second order reads%
\[
\omega \sigma _{3}u_{1}-\frac{1}{2}k^{2}u_{0}=-Lu_{2}.
\]%
Solvability of this equation leads to the dispersion law: 
\[
\omega \langle u_{0}|\sigma _{3}|u_{1}\rangle =\frac{k^{2}}{2}\langle
u_{0}|u_{0}\rangle 
\]%
or we have instability: 
\[
\omega ^{2}=-\frac{k^{2}\nu ^{2}}{3}<0.
\]%
Note that in small $\nu ^{2}=1-\kappa ^{2}<<1$, the growth rate of this
instability coincides with that found for the KDV soliton (compare with (\ref%
{KP-KDV})). The instability found represents a natural continuation of the
KP instability for gray (dark) solitons.  At the end of this subsection,  it is necessary to notice that the instability of dark solitons was confirmed
experimentally and numerically \cite{Schwartzlander}, \cite{Law}. In these
papers, in particular, it was demonstrated that the nonlinear stage of the
dark soliton instability results in  the formation of a point vortex street
familiar to the von Karman street in fluids (see also \cite{PelinovskyStepanyatsKivshar}). 

\subsection{Two-dimensional NLS solitons and their stability}

At the end of this section we consider two-dimensional solitons describing
within the NLSE (\ref{NLS}). 

In this subsection we will assume that $\psi \rightarrow 1$ at the infinity
for all directions. These solitons have a lot familiar properties with the 
lump solution. We restrict only  by  considering the axisymmetric soliton
solutions  $\psi _{0}(x^{\prime },y)=\psi _{0}(x^{\prime },-y)$ ($x^{\prime
}=x-Vt$). They represent stationary points of the Hamiltonian $\mathcal{H}$ for the
fixed total momentum $P$ which has only one nonzero $x$-projection:%
\begin{equation}
\delta (\mathcal{H}-VP)=0  \label{variation}
\end{equation}%
where 
\begin{equation}
\mathcal{H}=\frac{1}{2}\int \left[ |\nabla \psi |^{2}+(|\psi |^{2}-1)^{2}\right] dxdy,
\label{Ham}
\end{equation}
\begin{equation}
P=\frac{i}{2}\int \left[ \psi \frac{\partial \psi ^{\ast }}{\partial x}-\psi
^{\ast }\frac{\partial \psi }{\partial x}\right] dxdy.  \label{2Dmoment}
\end{equation}

Let $\varepsilon $ be the soliton energy (the Hamiltonian value on the
soliton solution) and $P$ is its momentum. Then, in accordance with (\ref%
{variation}),  the soliton velocity $V$  $V$ is written in terms of $\varepsilon $ and $P$ in the following form 
\begin{equation}
V=\frac{\partial \varepsilon }{\partial P}.  \label{velocity}
\end{equation}%
The variational problem allows also to express all integrals in (\ref{Ham}) 
 on the soliton solution in terms of  $\varepsilon $ and $P$ (\ref{2Dmoment}). 
For this aim, 
let us perform two independent scaling transformations along $x$ and $y$ axises: 
\[
\psi _{0}(x,y)\rightarrow \psi _{0}(ax,y),\psi _{0}(x,y)\rightarrow \psi
_{0}(x,by)
\]%
where $a$ and $b$ are scaling parameters. Hence, due to (\ref{variation}), it is easy 
to get two relations  
\[
\frac{\partial }{\partial a}(H-VP)|_{a=1}=\frac{\partial }{\partial b}%
(H-VP)|_{b=1}=0,
\]%
that gives the values of integrals on the soliton solution:
\begin{eqnarray*}
\int \left\vert \frac{\partial \psi }{\partial x}\right\vert ^{2}dxdy
&=&\varepsilon , \\
\int \left\vert \frac{\partial \psi }{\partial y}\right\vert ^{2}dxdy
&=&\varepsilon -VP, \\
\int (|\psi |^{2}-1)^{2}dxdy &=&VP.
\end{eqnarray*}%
Comparing with (\ref{velocity}), we get the inequality
\begin{equation} \label{inequality}
\frac{\varepsilon }{P}>\frac{\partial \varepsilon }{\partial P},
\end{equation}
which, according to the Bogolyubov spectrum (\ref{Bogolyubov}), has the opposite sign for linear waves,
\[
\frac{\omega }{k}<\frac{\partial \omega }{\partial k}.
\]%
 It is worth noting  once more that the soliton velocity  should be less the minimal
phase velocity of linear waves which, in our case, is the sound velocity ( $c_{s}=1$).

When the density (intensity) wells are small, but the condition $c_{s}-V>0$ is fulfilled, then the solution of this equation will approach the two-dimensional KP soliton (lump) and, accordingly, the equation (\ref{statNLS}) will turn into the stationary KP equation (\ref{stat-2D}). From this, one can immediately conclude that these two-dimensional solitons in the limit of shallow wells will be subject to KP instability with respect to modulations along the $z$ direction. For a larger soliton amplitude, this issue requires additional analysis.

An increase in the depth of these wells and, accordingly, a decrease in the soliton velocity $V$, as was first shown numerically in \cite{Jones}, occurs up to a certain critical value $V_{cr}$. For this critical velocity 
 the density at the center of the soliton vanishes. Below this critical value, the soliton solution undergoes a bifurcation associated with the appearance of two density zeros corresponding to two vortices with opposite circulations. This is a dipole pair propagating in a direction perpendicular to the dipole. With a further increase in the soliton velocity, this pair of zeros turns into a vortex pair for two-dimensional Euler hydrodynamics.

In this subsection we will show, following \cite{KuznetsovRasmussen},
that the entire family of two-dimensional solitons is unstable with respect to transverse (with respect to the $z$ axis) perturbations.
For a small soliton amplitude, this instability coincides with the KP
instability, in full agreement with the results of the previous subsections. For solitons propagating with velocities below the critical one, as the distance between the vortices grows, this instability transforms into the instability, discovered for the first time by Crow for the dipoles of two point vortices in the case of ideal incompressible fluids \cite{Crow}.

The growth rate in the long-wave limit for the whole family of the soliton
solutions is found by the same scheme presented above. For small
perturbations propagating with the velocity $V$ which in this frame depend
on  $z$ and $t$ exponentially,  $\sim \exp (-i\omega t+ikz)$, we
arrive at the following eigenvalue problem:
\begin{equation}
\omega \sigma _{3}u-\frac{1}{2}k^{2}u+Lu=0,  \label{3D-eigen}
\end{equation}%
where instead of the second derivative with respect to $x$ in the operator $L$ for the spectral problem (\ref{gray-stability})
stands the Laplace operator with respect to $x$
and $y$: 
\[
L=-i\kappa \sigma _{3}\frac{\partial }{\partial x}+\frac{1}{2}\Delta -\left( 
\begin{array}{cc}
2|\psi _{0}|^{2}-1 & \psi _{0}^{2} \\ 
\psi _{0}^{\ast 2} & 2|\psi _{0}|^{2}-1%
\end{array}%
\right) .
\]
In this case as before we will seek a solution of (\ref{3D-eigen}) in the form of power series relative to small $k,$
\[
u=u_{0}+u_{1}+u_{2}+...,\,\, \omega =\omega _{1}+\omega _{2}+...
\]%
In the zero approximation, $Lu_{0}=0$, we have two neutral modes
corresponding to two independent small shifts of the soliton, 
\[
u_{0}=C_{1}u_{01}+C_{2}u_{02}
\]%
where 
\[
u_{01}=\frac{\partial }{\partial x}\left( 
\begin{array}{c}
\psi _{0} \\ 
\psi _{0}^{\ast }%
\end{array}%
\right) ,u_{02}=\frac{\partial }{\partial y}\left( 
\begin{array}{c}
\psi _{0} \\ 
\psi _{0}^{\ast }%
\end{array}%
\right) 
\]%
and $C_{1}$ and $C_{2}$ are constants. 

The next order equation for the first mode corresponding to small shift $u_{01}$
\[
\omega \sigma _{3}u_{01}+Lu_{11}=0,  
\]
has a solution familiar to (\ref{1storder}), 
\[
u_{1}=i\omega \frac{\partial }{\partial V }\left( 
\begin{array}{c}
\psi _{0} \\ 
\psi _{0}^{\ast }%
\end{array}%
\right). 
\]
The procedure familiar to that in the previous subsection gives the answer for $\omega^2$, 
\[
\omega^2=\frac{\varepsilon}{\partial P/\partial V}k^2.
\]
The derivative $\partial P/\partial V$, as was shown analytically in \cite{KuznetsovRasmussen}, is negative in two limits, i.e. for the lump  and for solutions with small velocities $V\ll c_s=1$. In the first case 
\[
\omega^2=-2(1-V)k^2<0,
\]
which coincides with the the square of the instability growth rate for the lump. 
For small velocities $V\to 0$, the distance $L$ between vortices grows $\sim V^{-1}$, and the energy $\varepsilon \approx 2\pi \log (1/V)$. Then the definition of velocity (\ref{velocity}) implies that the derivative is negative: $\partial P/\partial V=-2\pi V^{-2}$. As a result, we get an instability with an growth rate $\gamma$, the square of which
\[
\gamma^2=2(kV)^2\log(1/V)>0.
\]
This growth rate coincides with that for the Crow instability  \cite{Crow} for two antiparallel point vortices.

In the general case, according to the numerical results \cite{Jones}, the derivative $\partial P/\partial V$ is negative for the entire family of two-dimensional solitons. 

Thus, the two instabilities, i.e., the KP and Crow instabilities, are in a certain sense analogous to each other. As the soliton velocity decreases, the KP instability transforms into the Crow instability, and therefore these instabilities should be combined. Both of these instabilities are of the self-focusing type. 

At the end of this section we would like to mention that the instability of the KP type can be derived also for 1D solitons \cite{DyachenkoKuznetsov} within the so-called Shrira equation \cite{Shrira} which describes the long-wave nonlinear perturbations in the boundary layer.

\section{Collapse}

As was shown in the second section, the Hamiltonian (\ref{H-KP}) for the three-dimensional KP equation with positive dispersion for a fixed $P$ is a functional unbounded from below, that follows from the scaling transformations for $\mathcal{H}(a,b)$ (\ref{H-scaling}). The unboundedness of $\mathcal{H}$ arises on small scales due to the nonlinear term. It is well known (see, for example, the review \cite{ZakharovKuznetsov2012}) that the unboundedness of the Hamiltonian for nonlinear wave systems is one of the main criteria for the occurrence of collapse. The collapse in this case can be understood as the process of falling particles in a self-consistent unbounded potential.

Note that under scaling along the parabolas $b \varpropto a ^{2}$ the first two
terms in (\ref{H-scaling}) behave like $a^{-2}$, and the cubic term responsible for the non-linearity changes
proportional to $a^{-5/2}$. For this reason, a possible collapse is not critical - it must be weak, as we will see below, it corresponds to a self-similar collapse. In such regimes, the radiation of low-amplitude waves from the collapse region promotes the collapse, which is again related to the unboundedness of $\mathcal{H}$. 

To show this, consider some area $\Omega $ with negative Hamiltonian, $\mathcal{H}_{\Omega }<0$. Using
mean value theorem, we can write an estimate for the maximum value $%
|u|_{\max}$ in this area:
\[
\int_{\Omega}u^{3}d{\bf r}\leq |u|_{\max}\int_{\Omega}u^{2}d{\bf r}.
\]%
Using this inequality under the condition $\mathcal{H}_{\Omega }<0$,
one can obtain the following estimate for $|u|_{\max}$ \cite{KuznetsovMusher1986}:
\begin{equation}
|u|_{\max}\geq \frac{|\mathcal{H}_{\Omega}|}{2P_{\Omega}}.
\label{radiation}
\end{equation}%
This shows that due to the radiation of waves from this region, the ratio on the right hand side of this inequality
will only grow over time. First, the radiated waves carry away the positive part of the Hamiltonian, since far from this region they become linear, for which the nonlinearity tends to zero. This leads to
decrease in $\mathcal{H}_{\Omega }$ (i.e., the Hamiltonian becomes more negative due to
unboundedness of $\mathcal{H}$) and, accordingly, an increase in $|
\mathcal{H}_{\Omega}|$. Second, at the same time $P_{\Omega }$, as a positive value,
decreases. It follows that the relation in (\ref{radiation})
 increases, and, consequently, $|u|_{\max}$ increases. This process,
similar to the evaporation of water droplets, leads to the formation of
weak singularities, when the energy $E$ of the acoustic waves captured in the singularity (coinciding with $P$ up to a constant factor) formally vanishes at the moment of collapse.

The numerical experiment \cite{KuznetsovMusherShafarenko1983, KuznetsovMusher1986} on simulating sonic collapse fully confirmed that radiation from the cavern region contributes to the collapse.
In these numerical experiments, for cylindrically symmetric distributions ($P_{\perp}=0$), zero boundary conditions were set for $u$ in $r$ and $x$, which, due to the nonlocality of the KP equation, do not preserve either $\mathcal{H }$ nor $P_x$. This made it possible to find out how radiation from the collapse region affects changes in the Hamiltonian, momentum, and maximum value of $u$.
\begin{figure}[tbp]
\begin{center}
\includegraphics[width=10cm,height=10cm,angle=0]{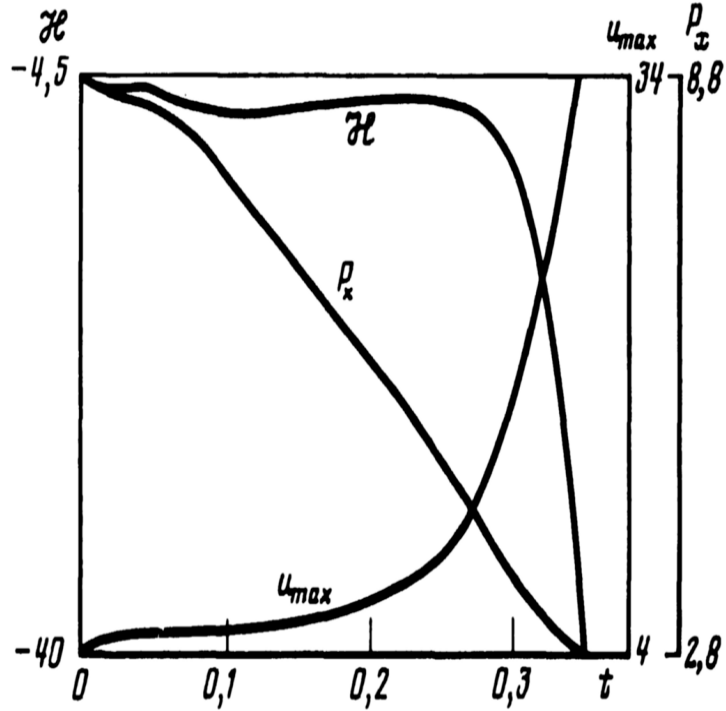}
\end{center}
\caption{Time dependences of the Hamiltonian $\mathcal{H}$, $P_x$ and maximum amplitude}
\end{figure}
The fig. shows the  time dependences of the Hamiltonian
$\mathcal{H}$ and $P_x$ computed over the integration domain and the maximum
amplitude. It can be seen that the momentum $P_x$ decreases with time and $|\mathcal{H}|$ increases with a simultaneous increase in $\max_{\mathbf{r}} u$, which just indicates that the collapse is weak. 

\subsection{Variational ansatz}

All the above considerations, however, do not answer the main question about
finiteness of the singularity formation time. We will bring
some arguments in favor of the finiteness of this time, using the so-called variational method. It should be noted that this approach is rather crude. For example, in the problem of relaxation of solitons in a one-dimensional NLSE with a focusing nonlinearity that can be integrated by the  inverse scattering transform, the application of the variational approach gives an incorrect answer. Instead of the soliton oscillations predicted by this method, the exact solution gives oscillations of the relaxation type damped with time \cite{KuznetsovMikhailovShimokhin}. However, in the collapse problem for the 3D NLSE, it turned out that the application of the variational approach demonstrated a qualitatively correct behavior for describing the strong collapse regime \cite{ZakharovKuznetsov1986}.

As is known. action for the KP equation (\ref{KP-stand}) can be
written as
\begin{equation}
S=\int \left[ \frac{1}{2}w_{t}w_{x}-\frac{1}{2}w_{xx}^{2}-\frac{1}{2}\left(
\nabla_{\perp}w\right)^{2}+w_{x}^{3}\right] dtdr, \label{action}
\end{equation}
so $\delta S=0$ is equivalent to the equation (\ref{KP-stand}). Let us choose the test function in the self-similar form
\begin{equation}
w=a^{1/2}b^{-1}f(\xi_{\parallel},\xi_{\perp}),\,\,u=a^{-1/2}b^{-1} U(\xi
_{\parallel},\xi_{\perp}), \label{test}
\end{equation}%
where $a$ and $b$ are assumed to be unknown functions of $t$, $%
\xi _{\parallel }=x/a$ and $\xi _{\perp}=r_{\perp }/b$ are self-similar variables. Function $U(\xi _{\parallel },\xi _{\perp })=\partial f/\partial \xi _{\parallel}$ is a solution  for a 
three-dimensional soliton of (\ref{KP-stand}) 
with known values of the integrals $I_{1}$, $I_{2}$ and $I_{3}$ determined by 
(\ref{1st}, \ref{2nd}, \ref{3d}): 
\[
I_{1}=6VP,\,\,I_{2}=I_{3}=4VP.  
\]
By substituting test function (\ref{test}) into (\ref{action}) and averaging over spatial
variables we get the following expression for the Lagrangian in terms of 
$a$ and $b$:
\[
L=a\frac{b_{t}}{b}M-\mathcal{H}( a,b). 
\]%
Here constant $M=\frac{1}{2}\int f\xi _{\perp }\nabla _{\xi _{\perp
}}Ud^{3}\xi $ and $\mathcal{H}\left( a,b\right) $ is already known function
given by (\ref{H-scaling}) at $d=3$, which is obtained by means of the scaling transformations.

For this Lagrangian the Euler-Lagrange equations have the form%
\begin{eqnarray}
Ma_{t} &=&-b\frac{\partial \mathcal{H}}{\partial b}%
=a^{2}b^{-2}I_{2}-a^{-1/2}b^{-1}I_{3},  \label{alpha} \\
Mb_{t} &=&b\frac{\partial \mathcal{H}}{\partial a}=\beta \left[
-a^{-3}I_{1}+ab^{-2}I_{2}+\frac{1}{2}a^{-3/2}b^{-1}I_{3}\right] .
\label{beta}
\end{eqnarray}%
These equations, by construction, have a stationary solution in the form of 3D solitons when $a=b=1$. Linear stability of this solution is defined from the linearized equations (\ref{alpha}, \ref{beta}) for small perturbations $\alpha ,\beta $
($a=1+\alpha $, $b=1+\beta $):
\begin{eqnarray*}
M\alpha _{t} &=&VP\left( 10\alpha -4\beta \right) , \\
M\beta _{t} &=&VP\left( 19\alpha -10\beta \right) .
\end{eqnarray*}%
Hence, we arrive at the instability independently on the sign of $M$ with
the following growth rate%
\[
\gamma =\pm 2\sqrt{6}\frac{VP}{M}. 
\]
The eigen vector for the unstable mode has the same sign for both $\alpha$
and $\beta $ components. Thus, the soliton solution is a saddle
point.

Now, consider what will happen at the nonlinear stage of this instability.
As mentioned above, at the fixed momentum $P$ the Hamiltonian of the 3D KP
equation represents the unbounded functional from below. 
In the given case we deal with  unbounded from below function $\mathcal{H}( a,b) $.

Since the system (\ref{alpha}, \ref{beta}) is Hamiltonian, it conserves energy  $E=\mathcal{H}\left( a,b\right) $ that permits to express $b$ through $a$,
\[
b^{-1}=\frac{1}{I_{2}a^{2}}\left[ a^{-1/2}I_{3}\pm \sqrt{%
a^{-1}I_{3}^{2}-I_{1}I_{2}+2a^{2}EI_{2}}\right] , 
\]%
and then, after excluding $b$, to define the temporal dependence $a(t)$
near singularity for both roots. For $a\rightarrow 0$ 
\[
b_{1}=\frac{I_{2}}{2I_{3}}a^{5/2},b_{2}=\frac{2I_{3}}{I_{1}}a^{3/2}. 
\]%
Hence for the first root we have%
\[
Ma_{t}=\frac{2I_{3}^{2}}{I_{2}a^{3}}. 
\]%
 This leads to the singular solution
\begin{equation}
a\rightarrow \left( t_{0}-t\right) ^{1/4},  \label{first-root}
\end{equation}%
if constant $M<0$.
In this case, near the collapse moment $t=t_{0}$ in $\mathcal{H}( a,b) $, the diffraction and nonlinear terms cancel each other out. For positive $M$ there is no solution with a singularity.

For another root $b=b_{2}$ for $a\rightarrow 0$ the equation
(\ref{alpha}) asymptotically transforms to the form
\[
Ma_{t}=-\frac{I_{1}}{2a^{2}},
\]
which for $M>0$ has a collapsing solution %
\[
a\rightarrow \left( t_{0}-t\right) ^{1/3}.  
\]
For this asymptotics the nonlinear term and the dispersion one in $\mathcal{H%
}( a,b) $ are canceled. It will be shown below that the mode (\ref{first-root}) is implemented for  the quasi-classical initial conditions.

\subsection{Quasi-classical collapse}

Consider the initial distributions $u$ for which both dispersion and diffraction can be neglected. Then, at least initially, the temporal behavior of $u$ will be determined by the equation%
\[
u_{t}+6uu_{x}=0,
\]
where $u$ depends on ${\bf r}_{\perp }$ as a parameter. As is known, this equation describes wave breaking or gradient catastrophe that occurs at some point ${\bf r}={\bf r}_{0}$. Near the breaking point, our assumption about dispersion and diffraction becomes invalid, and we must take into account both diffraction and dispersion terms. As a result, due to these linear effects, an oscillating spatial structure will develop near this point. Recall that in the one-dimensional case of such kind structure has the form of a spreading soliton train \cite{GurevichPitaevskii}, which is characteristic for collisionless shock waves.

In the three-dimensional case, as we found out, all types of solitons are unstable. One-dimensional solitons
unstable with respect to transverse modulations due to the KP instability. The same statement is true for two-dimensional solitons \cite{KuznetsovTuritsyn1982}. The nonlinear stage of this instability must be a collapse. This means that instead of a train of solitons in one-dimensional geometry, one should expect the formation of oscillating structures containing collapsing solitons. 

To describe structures of this type, it is necessary to turn to the 
quasiclassical theory. The solution in this case should be presented as %
\[
u=u(r,t,\Phi ({\bf r},t))
\]%
where $u$, as a function of the phase $\Phi$, is $2\pi$-periodic and, with respect to the first two arguments $r$ and $t$, a slowly varying function.
  The dependence $\Phi(r,t)$ is assumed to be such that its derivatives $\Phi _{t}$ and $\nabla \Phi $ are also slowly varying functions of ${\bf r}$ and $t$. 

First, let us consider the linear case when $u$ obeys the linear KP equation%
\begin{equation}
\frac{\partial }{\partial x}\left( u_{t}+u_{xxx}\right) =\Delta _{\perp }u.
\label{linearKP}
\end{equation}%
In this situation, we can restrict ourselves by a one-harmonic dependence%
\[
u=Ae^{i\Phi }+c.c.. 
\]%
Then, in the quasiclassical approximation, the first order leads to the Hamilton-Jacobi equation 
\begin{equation}
\Phi _{t}+\omega (\nabla \Phi )=0,  \label{linearHJ}
\end{equation}%
where $\omega =\omega (k)=-k_{x}^{3}-k_{\perp }^{2}/k_{x}$ is the dispersion
law for (\ref{linearKP}) and $k=\nabla \Phi $ the wave vector, respectively.

In the next order, we get the continuity equation for $A^{2}$ (here for
simplicity $A$ is supposed to be real),%
\begin{equation}
\frac{\partial A}{\partial t}+\frac{1}{2A}\mbox{div}(A^{2}\mathbf{v})=0,
\label{linear-cont}
\end{equation}%
where $\mathbf{v}=\partial \omega (\mathbf{k})/\partial \mathbf{k}$ is the
group velocity (compare with \cite{ZakharovKuznetsov1986, KuznetsovTuritsyn1990}). Because of the nonlinearity of the original equation (\ref%
{KP-stand}) we have to take into account all harmonics,%
\begin{eqnarray*}
u &=&\sum_{n=-\infty }^{\infty }A_{n}e^{in\Phi }, \\
A_{n} &=&A_{-n}=A_{n}^{\ast }.
\end{eqnarray*}%
As a result,  instead of (\ref{linear-cont}) and (\ref{linearHJ}), we have
for $n\neq 0$%
\begin{eqnarray}
\frac{\partial A_{n}}{\partial t}+\frac{1}{2A_{n}}\mbox{div}(A_{n}^{2}%
\mathbf{v}_{n})+3\frac{\partial }{\partial x}s_{n} &=&0,  \label{cont} \\
\Phi _{nt}+\omega (\nabla \Phi _{n})+3\Phi _{nx}\frac{s_{n}}{A_{n}} &=&0,
\label{HJ}
\end{eqnarray}%
where $\Phi _{n}=n\Phi $ and $s_{n}=\sum_{n=n_{1}+n_{2}}A_{n_{1}}A_{n_{2}}$.
For zero harmonics a separate equation arises, 
\begin{equation}
\frac{\partial A_{0}}{\partial t}+3\frac{\partial }{\partial x}\langle
u^{2}\rangle =0.  \label{zero}
\end{equation}%
Here 
\[
\langle u^{2}\rangle =\sum_{n=-\infty }^{\infty }A_{n}^{2}=s_{0}. 
\]%
It should be noted that this infinite system is overdetermined because from (%
\ref{cont}, \ref{HJ}) for each $n,m\neq 0$ we get the following constraints%
\begin{equation}
\pi _{n}(r,t)=\pi _{m}(r.t)  \label{constraints}
\end{equation}%
where 
\[
\pi _{n}(r,t)=n^{2}\Phi _{x}^{2}-\frac{s_{n}}{A_{n}}. 
\]%
At first glance, these equations seem to be very complicated and
unlikely to be analyzed. In fact, this system of equations is completely equivalent to the equations that follow from Whitham's \cite{Whitham, GurevichPitaevskii} procedure for the KDV case. First of all, it is easy to find the number of independent functions given all
constraints (\ref{constraints}). These are just three functions that we can, for example, choose $\langle u\rangle $, $\langle
u^{2}\rangle $ and $\Phi $. Moreover, the conditions (\ref{constraints}) can be resolved if $u$ is represented as a cnoidal wave using the Weierstrass elliptic function (see, for example, \cite{KuznetsovMikhailov1974}). As a result, we can get the usual form for the Whitham equations. However, in our opinion, the equations (\ref{cont}), (\ref{zero}) for the study of quasiclassical collapse seem to us more convenient than the Whitham equations.  

\subsection{\protect\bigskip Collapse Hierarchy}

\bigskip 
Let us seek for the solution of system (\ref{cont} -\ref{zero}) in
the self-similar form%
\begin{eqnarray}
A_{n}(r,t) &=&(t_{0}-t)^{-\alpha }f_{n}\left( \frac{x}{(t_{0}-t)^{\beta }},%
\frac{r_{\perp }}{(t_{0}-t)^{\gamma }}\right) ,  \label{ampl} \\
\Phi (r,t) &=&\lambda ^{2}\int^{t}\frac{dt}{(t_{0}-t)^{\kappa +1}}%
+(t_{0}-t)^{-\kappa }\varphi \left( \frac{x}{(t_{0}-t)^{\beta }},\frac{%
r_{\perp }}{(t_{0}-t)^{\gamma }}\right)  \label{phase}
\end{eqnarray}%
where $t_{0}$ is a collapse time, $\lambda ^{2}$ is a constant, $\alpha
,\beta ,\gamma ,$ and $\kappa $ are unknown indices.

After substitution (\ref{ampl}), (\ref{phase}) into system (\ref{cont} -\ref%
{zero}) it is easy to find out that only one index is free (let it will be $%
\alpha $) and the others are expressed through it%
\[
\beta =1-\alpha,\,\,\gamma =1-\alpha /2,\,\,\kappa =3\alpha /2-1. 
\]%
Thus, we have the whole family of the self-similar solutions (more
rigorously, substitutions). Only some of them can have physical meaning.
According to (\ref{ampl}), (\ref{phase}), the longitudinal size of the
collapsing region (cavity) and the transverse one are proportional to $%
(t_{0}-t)^{\beta }$ and $(t_{0}-t)^{\gamma }$, respectively. 
 The physical requirement is that the complete
cavity momentum $P$ (actually, this is the energy of sound waves up to a constant
multiplier) during the collapse due to radiation can only decrease:
\[
P_{cav}=\frac{1}{2}\int_{cav}u^{2}d^{3}r\propto (t_{0}-t)^{3-4\alpha }, 
\]%
i.e. $\alpha \leq 3/4$. This shows that $\alpha =3/4$ will correspond to
 strong collapse regime. It should be noted that the asymptotics for this type
of collapses is the same as for the variational ansatz (\ref{first-root}). This gives reason to believe that such a regime may exist. All other collapses are weak.

For the self-similar family (\ref{ampl}), (\ref{phase}), its lower boundary is determined based on the quasiclassical criterion. The latter means that the ratio between the characteristic period, $\propto\omega ^{-1}=\left( \partial_{t}\Phi \right) ^{-1}$, and the characteristic time of amplitude change $ \propto (t_ {0}-t)$, or the ratio of the wavelength $k_{x}^{-1}=\left( \partial _{x}\Phi \right) ^{-1}$ to the characteristic
spatial size of the amplitude must be large, $\gg 1$. Both of these relationships
lead to the same criterion:
\begin{eqnarray*}
&&\omega (t_{0}-t)\propto (t_{0}-t)^{1-3\alpha /2}, \\
&&k_{x}(t_{0}-t)^{1-\alpha }\propto(t_{0}-t)^{1-3\alpha /2}.
\end{eqnarray*}%
From here it can be seen that for $\alpha >2/3$ the quasiclassical criterion improves while approach the collapse point. $\alpha =2/3$ represents the lower boundary of this family, where the quasiclassical criterion fails. The value $\alpha =2/3$ corresponds to a self-similar solution of the three-dimensional KP equation, which describes the fastest weak collapse. 

\section{Conclusion}
Thus, we have demonstrated the possibility of the collapse of acoustic waves in media
 with positive dispersion based on the three-dimensional
Kadomtsev-Petviashvili equation. This process can be viewed as
nonlinear stage of the KP instability of KDV solitons in the transverse
direction. This instability has a simple explanation, first given by B.B. Kadomtsev \cite{kadomtsev1976collective}.
Solitons with positive dispersion are regions of reduced density, the velocity of which decreases with increasing soliton amplitude.
For this reason, when modulating a soliton 
in the transverse direction to its propagation, regions with a lower amplitude overtake regions with a higher amplitude. As a result, the soliton front experiences self-focusing instability, which leads to the accumulation of the soliton distribution.
This fact is a key point for understanding the nature of the KP instability. It's important that
this character of the instability is retained at its nonlinear stage, which leads to collapse in the three-dimensional case. At lower dimensions, for example, in two dimensions, the soliton (lump) of the KP  equation realizes the minimum of the Hamiltonian at a fixed momentum $P$ and, for this reason, turns out to be stable in the Lyapunov sense. In three-dimensional geometry, solitons are saddle points of the Hamiltonian and, therefore, turn out to be unstable. The development of this instability leads to collapse, one of the criteria for which is  the Hamiltonian $\cal{H}$  unboundedness from below for a fixed momentum $P$. The collapse in this case can be understood as the process of falling particle in a self-consistent potential unbounded from below.
Due to the unboundedness of the Hamiltonian, the formation of a singularity is possible due to the emission of low-amplitude waves from regions 
with ${\cal H}<0$, that promotes the collapse.
In this case, the radiation leads to the formation of a weak collapse, when the wave energy (coinciding with the $L_2$-norm up to a factor) captured into the singularity  formally tends to zero as it approaches the singularity point. It is shown that in the quasiclassical description of the collapse of acoustic waves with positive dispersion, a whole family of self-similar collapsing regimes is possible, starting from strong, when the energy captured into the singularity is finite, up to weak collapses accompanied by wave radiation. The lower boundary of this family corresponds to
self-similar solution of the three-dimensional KP equation, which describes the fastest of the weak collapses. The tendency for ${\cal H}$ to decrease upon collapse due to radiation was first demonstrated in numerical simulations \cite{KuznetsovMusherShafarenko1983, KuznetsovMusher1986}. In these numerical experiments
it was shown that the Hamiltonian of the cavity becomes
more and more negative over time, at the same time the value of $P$ decreases, that leads to the growth of the amplitude in the collapsing region.

Despite the predictions of the hierarchy of quasiclassical collapses, the role of radiation in these processes, etc., there are many
open questions. From our point of view, the most important is
a question regarding the exact collapse criterion for the 3D KP equation (\ref{KP-stand}) of the same level of rigor as the Vlasov-Petrishchev-Talanov criterion for the 2D focusing NLSE \cite{VlasovPetrishchevTalanov} or its generalization for the 3D case found by
Zakharov \cite{Zakharov1972} (for other generalizations, see \cite{KTRR}). These criteria provide sufficient
collapse condition for initial distributions with negative
Hamiltonian. It should be especially noted that the criterion for the three-dimensional KP equation, but with a cubic nonlinearity of the focusing type, was obtained by Turitsyn and Falkovich \cite%
{TuritsynFalkovich}. According to this work, a sufficient condition for
collapse is also the negativeness of the Hamiltonian.

Another open question concerns the construction of an analytical solution for the three-dimensional KP equation. Recall that, as was shown in \cite{Dryuma}, the two-dimensional KP equation admits the Lax representation and, accordingly, the application of the inverse scattering transform, the use of which has shown great efficiency in analytical studies of the two-dimensional KP equation. It suffices here to mention the work \cite{manakov1977twodimensional}, in which a solution in the form of a lump was constructed and the problem of scattering of two-dimensional solitons was studied. As is known, there is
another approach, the so-called Hirota  method \cite{Hirota}, which allows one to construct some analytical solutions for a large number of integrable models. Relatively recently
it has been demonstrated that the 3D KP equation also admits a Hirota representation
\cite{Ma}, \cite{Qian}. However, so far all built
3D KP solutions based on this approach turn out to be only 2D, see 
also \cite{Wang}, \cite{Mao}, \cite{Zhang}, \cite{Kaptsov2021}. 
Therefore, one of the main tasks is
the problem of constructing exact three-dimensional solutions that have a physical meaning. It must be emphasized that, strictly speaking, the presence of a Hirota representation does not guarantee the existence of a Lax representation. 

\section{Acknowledgment}
The author thanks A.M. Kamchatnov and Yu.A. Stepanyants for useful discussions and remarks.

\end{document}